# Tourism's Trend Ranking on Social Media Data using Fuzzy-AHP vs. AHP

Shoffan Saifullah
Department of Informatics Engineering, Universitas Pembangunan Nasional Veteran Yogyakarta
Jl. Babarsari 2 Yogyakarta
e-mail: shoffans@upnyk.ac.id

**ABSTRACT**

Tourism is an exciting thing to be visited by people in the world. Search for attractive and popular places can be done through social media. Data from social media or websites can be used as a reference to find current travel trends and get information about reviews, stories, likes, forums, blogs, and feedback from a place. However, if the search is done manually one by one, it takes a long time, and it becomes interesting to do research. So, searching based on current trends will be easier and faster. For this reason, this study uses a computer base to search by ranking tourist facilities from social media data or websites using the multi-criteria decision-making (MCDM) method. The implementation of the method used in finding the trend is the Fuzzy-AHP method in comparison with the AHP. The data used is data reviews, stories, likes, forums, blogs, and feedback from the web or social media. Because with these components, tourism can be developed according to visitors' wishes. The research aims to rank facilities' tourism attractions (trends) and development priorities. The priority and ranking used the fuzzy-AHP and AHP method to determine weight criteria and the ranking process. The highest ranking is on the Parks/Picnic Spots attraction, and make it a priority to develop. The methods have an average value MSE of all data is ≈ 0.0002, which can be used for this ranking.

## I. INTRODUCTION

Tourism is the largest sector in the world, many of which have many visitors [1]. Many people look for places that have uniqueness and privileges in visiting tourism. In addition, they look for tourist attractions with quality service and the site itself, as well as the feelings of people who have seen the tour [2], which are usually shared through social media. Therefore, tourism places usually become businesses for providers of products and services that have distinctive characteristics [3]–[5]. Based on this, each region has the potential to develop tourist areas [6] by involving the existing resources in the area; one example is a tourist attraction in East Asia [7]. In addition to existing resources, each destination will develop based on an approach to tourists/visitors, both from their activities, interests, and opinions [8]. So that tourists will visit again if the place has adequate facilities and becomes information to be shared with the public [6]. The information becomes a reference in the development and visits of other tourists. The application of technology can accelerate the dissemination of information via the internet, either through websites or social media [9]. Much internet is used nowadays; almost everyone can access it to share and invite others to travel to certain places through social media [10]. It can be a consideration for someone to visit tourism. The information content shared on social media can be used to increase the tourism potential of a place and can be used for rating and rating existing tourism and can be used to improve development priorities carried out.

Tourist visits are an attraction in conducting travel, tourism, and research activities. Many studies have carried out research in the field of tourism, such as computational approaches to specific tourism [7], [11]–[13], such as the evaluation of tourism websites [14]–[16]. In addition, it also carries out the process of ranking and ranking tourism. When the order is done manually, it will take a long time and analysis, so it requires an approach in the automated ranking process. One of the implementations is carried out in East Asia. The ranking method can be done using the Analytical Hierarchy Process (AHP) method, and its development is Fuzzy AHP. Things that become a reference in research

Tourism development is often done by observing tourists' responses and social media postings. Thus, social media can be used as a reference to see reviews of tourism assessments in East Asia, both for domestic and foreign tourists. In East Asia, developers take several approaches to determine how to prioritize the development of facilities based on their interest in tourism trends, one of which is by reviewing their tours on social media. Priority for selecting tourist attractions has been developed using the MCDM approach (TOPSIS and Fuzzy-AHP) based on tourist preferences [17]. Smart Tourist Attraction (STA) has also been designed to evaluate tourists using the Fuzzy AHP method and Importance-Performance Analysis (IPA) [18]. This is only limited to the evaluation process and its development recommendations. Several have implemented intelligent systems such as AHP and case-based reasoning (CBR) methods in developing tourist attractions planning applications. This system decides the plan based on the response from the customer. However, this system does not yet have clear filtering rules and preferences for changing customer responses. Thus, this study recommends fuzzy to resolve any possible uncertainties [19], [20].

This article contains several parts; the second part describes previous tourism research. It was then continued with the explanation of methods, data, and tools in the third part. The fourth section describes the results and discusses the experiments and their tests. Finally, explain the conclusions of this study.

## II. METHODS

This study uses a dataset from UCI Machine Learning. The dataset contains review data for tourist attractions in East Asia. The review consists of 10 destination category data from 980 reviews of social media user data (TripAdvisior.com). The review consists of 5 assessment categories: terrible, poor, average, good, and excellent. The assessment criteria range from 0 to 4, which in detail are shown in Table 1. Each rater is a traveler who gives an assessment categorized as 0 = terrible, poor ≤ 1, average ≤ 2, very good ≤ 3, and excellent ≤ 4. The assessment results will be a reference for ranking and rating each destination visited by tourists.

TABLE I
RANGE OF TOURIST RATING CATEGORIES FOR TOURIST DESTINATIONS

| No | Categories | Range |
|----|------------|-------|
| 1  | Terrible   | 0     |
| 2  | Poor       | 0,01-1 |
| 3  | Average    | 1,01-2 |
| 4  | Very Good  | 2,01-3 |
| 5  | Excellent  | 3,01-4 |

Data collection showed that the AHP and fuzzy AHP methods rank the average results, as shown in the flowchart in Figure 1. The first step is to evaluate average tourist data based on social media users. The data are in the form of matrices and are normalized to obtain small values and maximize computation. This normalization aims to align the attribute values and is performed using equation (1).

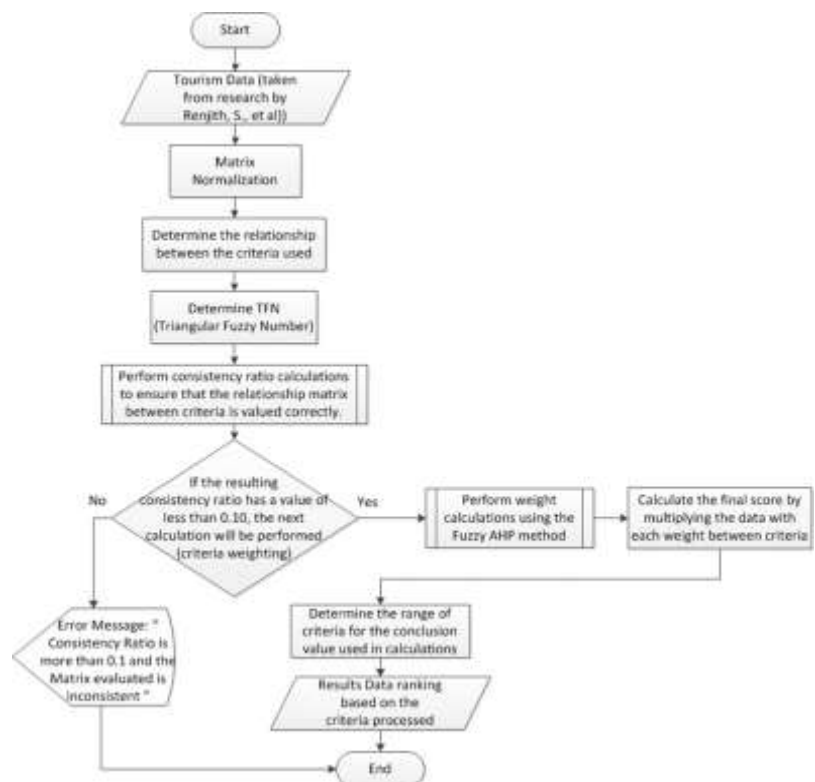

Figure 1. Flowchart for ranking tourist attractions from social media data use the AHP and Fuzzy AHP methods.

$$r_{ii} = \frac{x_{ij}}{max_i \, x_{ij}}; \; j = 1,2,3,\ldots,m; i = 1,2,3,\ldots,n \tag{1}$$

$r_{ij}$ is a normalized matrix, $x_{ij}$ is a decision matrix, and $max_i$ is the maximum value of each column. This normalization process uses the simple addictive weight (SAW) normalization equation [21].

A normalization matrix is used to determine the relationship between criteria. Relationships between criteria are established using Analytical Hierarchy Process (AHP) methods. AHP is used to find ratio scales based on pairwise comparisons of criteria [22]. AHP can handle quantitative and qualitative decision-making problems for simple problems [23]. The comparison scale used in the AHP method uses the rating scale proposed by Saaty and is divided into a range of 1 to 9 points.

AHP has been developed in several studies, one of which is a combination of fuzzy and AHP [24]. Fuzzy AHP is a method for developing AHPs that can handle fuzzy decisions [25][26]. The fuzzy used is TFN (Triangular Fuzzy Number), a set of three numbers that form a fuzzy graph with a fuzzy value of 0, increasing to 1 and back to 0 [27]. TFN contains two data groups. The first group is the TFN of the actual value, and the second group is the inverse of his TFN, i.e., change x to 1/x and reverse the order of the TFN numbers.

In the AHP method, decision-making makes it easier for people to make decisions. It is because humans use perception to tolerate contradictions when making decisions. AHP determines whether perception is consistent. This notion of consistency is formulated in the Consistency Index.

$$CI = \frac{t-n}{n-1} \tag{2}$$

CI is the consistency index, t is the top normalized value of ordered matrix n, and n is the ordered matrix. This AI calculation checks the integrity of Saaty's pairwise comparison matrix. The matrix is said to be consistent when the value of CI is zero (0). Saaty also explained that there are limits to inconsistencies in using consistency ratios.

$$CR = \frac{CI}{IR} \tag{3}$$

The Consistency Ratio (CR) is a comparison between the consistency index (CI) with a random index value (IR). The maximum CR limit is 0.1 (CR ≤ 0.1). This condition states that inconsistencies in decision-making are still accepted and will be reprocessed if not suitable. Acceptance of the decision-making process based on CR values will determine the weight calculation process. The method used in this calculation is Fuzzy AHP. The process includes 2 main steps: converting each relation between criteria into a Triangular Fuzzy Number (TFN) and calculating the degree of probability.

The initial process of Fuzzy AHP calculates the relation matrix between criteria by entering the value in the lower triangle according to the value in the relationship between criteria. Suppose the relation matrix value between criteria is more than 1. In that case, the TFN value used is the criterion value in the first group, in addition to using the criterion value in the second group. Moreover, the values of each TFN used in the matrix between criteria are added. The results of which are used to add values to each criterion. Besides, the calculation of each value is divided by the number of relations in each column. The estimation of the weight value can be obtained by calculating the degree of probability. This process considers the principle of comparison between fuzzy numbers. This comparison is obtained by determining the value of the vector (V) [28].

$$V(M2 \geq M1) = \begin{cases} 1 & , if \; m_2 \geq m_1 \\ 0 & , If \; l_1 \geq u_2 \\ \frac{(l_1 - u_2)}{(m_2 - u_2) - (m_1 - l_1)} & , other \end{cases} \tag{4}$$

Possible orders are obtained based on calculations for each data using row and column indices. The result is converted to the vector expression (3). The minimum possible grade value for each criterion is weighted and normalized by dividing each weighted value by all of the resulting weights. Finally, a list of criteria for the final calculation and ranking of all data is defined.

The final result of the normalized weights is computed as the mean squared error value that determines the accuracy of the ranking method. The Mean Squared Error (MSE) calculation represents the value of the squared error.

$$MSE = \frac{1}{n}\sum_{i=1}^{n}(f_i - y_i)^2 \qquad (5)$$

### III. RESULT AND DISCUSSION

Based on the experiment, this section discusses the dataset used in this research and the process. Moreover, this part presents the result of the methods (AHP and Fuzzy AHP) for ranking tourism destinations in East Asia. TFN method and criteria calculate for the ranked methods.

*A. Analysis of Dataset based on Distribution Value*

The form of the dataset in this research is a matrix with a size of 10 x 980 of data [29]. The dataset has 2 distributions, namely actual data and normalization data distribution. Based on the normalization, the process needs to be converted into data with a spread of 0 until 1, and it can be created using equation (1). The dataset can be seen by visualizing a line graph in Figure 2.

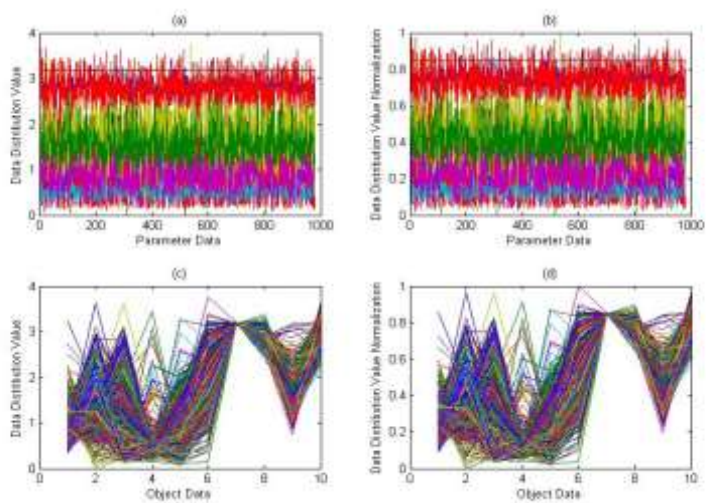

Figure 2. Dataset distribution based on East Esia tourism destination analyze on social media using parameters wtih the (a, c) real object and (b, d) normalized value showed by graph visualization

The tourism destination from the East Asia dataset, the initial data, and parameters are ultimately shown in Figures 2(a) and 2(c) with the actual data on their distribution. Based on this dataset, the overall data of this distribution should be normalized and get the distribution number with rang 0 to 1, such as in Figure 2(b) related normalization from 2(a) and Figure 2(d) related normalization from 2(c). The normalized data will be easy to use in the following process, especially on Fuzzy AHP and AHP calculations.

*B. Triangular Fuzzy Number (TFN) and criteria*

Parameter data are used to calculate the relationships between the criteria using TFN. The process should use the number of TFN (Table 2), and this number has 9 label characteristics, from "just equal" to "extremely strong." In addition, this criterion has real and inverse numbers with 3 members of the TFN sum. The intensity of AHP's importance shows that TFN members use numbers with values from 1 to 9 divided by 2 for each label.

TABLE II
COMPARATIVE SCALE OF CRITERIA USING TFN

| Intensity of Interest AHP | Linguistic labels | TFN Real | TFN Inverse |
|---|---|---|---|
| 1. | Just Equal | 1 1 1 | 1 1 1 |
| 2. | Intermediate | $\frac{1}{2}\ \frac{3}{4}\ 1$ | $1\ \frac{4}{3}\ 2$ |
| 3. | Moderately Important | $\frac{2}{3}\ 1\ \frac{3}{2}$ | $\frac{2}{3}\ 1\ \frac{3}{2}$ |
| 4. | Intermediate | $1\ \frac{3}{2}\ 2$ | $\frac{1}{2}\ \frac{2}{3}\ 1$ |
| 5. | Strong Important | $\frac{3}{2}\ 2\ \frac{5}{2}$ | $\frac{2}{5}\ \frac{1}{2}\ \frac{2}{3}$ |
| 6. | Intermediate | $2\ \frac{5}{2}\ 3$ | $\frac{1}{3}\ \frac{2}{5}\ \frac{1}{2}$ |
| 7. | Very Strong | $\frac{5}{2}\ 3\ \frac{7}{2}$ | $\frac{2}{7}\ \frac{1}{3}\ \frac{2}{5}$ |
| 8. | Intermediate | $3\ \frac{7}{2}\ 4$ | $\frac{1}{4}\ \frac{2}{7}\ \frac{1}{3}$ |
| 9. | Extremely Strong | $\frac{7}{2}\ 4\ \frac{9}{2}$ | $\frac{2}{9}\ \frac{1}{4}\ \frac{2}{7}$ |

Based on Table 2, the intensity of the interest in AHP has a relationship between the criteria and the TFN values. The TFN and inverse members should know that the number is the opposite of the actual members. For example, based on label 2, the actual number of TFN is $\frac{1}{2}\ \frac{3}{4}\ 1$, and the inverse should be $1\ \frac{4}{3}\ 2$. It means that the number from the first position should be changed in the third number and reversed.

If the calculation of the consistency ratio (CR) of the AHP calculation is <0.1, the process should use the TFN calculations. This research process used several variables on AHP dan Fuzzy AHP consistency calculation like random index, lambda values, consistency index, and consistency ratio. The experiments have shown the result parameters (Table 3) used on the AHP calculation and Fuzzy AHP.

TABLE III
DATA AND RESULTS OF AHP CONSISTENCY ON AHP AND FUZZY AHP

| Result Variable | Fuzzy AHP | AHP |
|---|---|---|
| Random Index | range 0 s/d 180 | - |
| Lambda Max. | 1 | 4.37 |
| Average weight between criteria | 0.001 | - |
| Consistency Index | -1 | 0.126 |
| Consistency Ratio | -0.0056 | 0.228 |

The random range of the index affects the consistency index (CI) and the calculation of the consistency ratio (CR). The higher the random index range, the resultant consistency ratio is close to 0 (zero). Table 3 shows that the CR value of Fuzzy AHP is -0.0056, which means that the CR condition is less than 0.1 and will be processed by quantified fuzzy AHP. Based on this process, TFN ness some requirement components, like degree of probability, the number of relations, and weight maximum (shown in Table 4). Besides that, the CR of AHP is 0.2284, which means that the AHP should be calculated and used the repair the number of CR until normal and will be calculated with AHP processed. The AHP process will be initiated by paired matrix (of the criteria and calculated by eigenvector.

TABLE IV
THE VALUE OF VARIABLES OF FUZZY AHP AND PAIRED MATRIX OF AHP

| Fuzzy-AHP Variables | Values | Paired Matrix of AHP | | | |
|---|---|---|---|---|---|
| | | 1,00 | 1,33 | 2,00 | 4,00 |
| Avg degree of prob. | 327.33 | 0,75 | 1,00 | 1,50 | 3,00 |
| Avg numb. of relations | 0.001 | 0,50 | 0,67 | 1,00 | 2,00 |
| Weight Max. | 1 | 0,25 | 0,33 | 1,33 | 1,00 |

Based on these parameters, the results produced output weight values. Fuzzy AHP and AHP calculate the weight result and data by multiplication, and the data will be ranked based on the parameters.

*C. The result of destination ranking using Fuzzy AHP vs. AHP*

Based on the previous calculation, the results are shown in Figure 3 for processing each dataset and predefining the dataset (manual and normalization). The ranking results on each scenario (AHP, Fuzzy-AHP, and manual calculation) are shown in Figure 3 from largest to smallest (descending). Based on the experiment, the results show that the parks/picnic spots are the first place on the rank with a significant value of 0.4 on AHP, 0.6 on Fuzzy AHP, and 0.8 on Manual Normalization. Detailed results of these calculation methods and the ranks for each tourist destination are shown in Table 5.

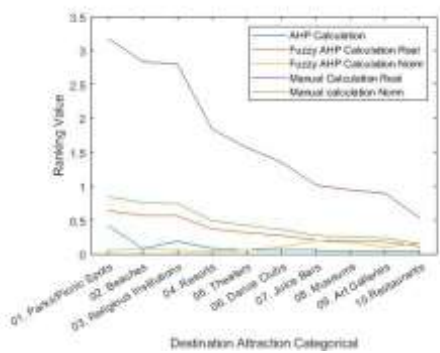

Figure 3. The results of AHP vs Fuzzy-AHP vs Manual calculation results at each tourist site based on users' feedback.

TABLE V
COMPARISON OF THE TOURIST ATTRACTIONS RANK BY AHP VS FUZZY-AHP METHOD VS MANUAL CALCULATION ON SOCIAL MEDIA DATA

| Rank | Tourist Attractions | Average user feedback on and the results | | | | |
|---|---|---|---|---|---|---|
| | | AHP | Fuzzy AHP | | Manual | |
| | | | $1^a$ | $2^b$ | $1^a$ | $2^b$ |
| 1 | Parks/Picnic Spots | 0.4207 | 0.6361 | 0.0555 | 3.1809 | 0.8459 |
| 2 | Beaches | 0.0790 | 0.5670 | 0.0743 | 2.8350 | 0.7540 |
| 3 | Religious Institutions | 0.1842 | 0.5598 | 0.0560 | 2.7992 | 0.7444 |
| 4 | Resorts | 0.0790 | 0.3685 | 0.0310 | 1.8428 | 0.4901 |
| 5 | Theaters | 0.0525 | 0.3138 | 0.0570 | 1.5694 | 0.4174 |
| 6 | Dance Clubs | 0.0525 | 0.2705 | 0.0980 | 1.3526 | 0.3597 |
| 7 | Juice Bars | 0.0525 | 0.2026 | 0.1982 | 1.0133 | 0.2694 |
| 8 | Museums | 0.0264 | 0.1879 | 0.1673 | 0.9397 | 0.2499 |
| 9 | Art Galleries | 0.0264 | 0.1786 | 0.0990 | 0.8931 | 0.2375 |
| 10 | Restaurants | 0.0264 | 0.1065 | 0.1530 | 0.5325 | 0.1416 |

[a.] Real value calculation
[b.] Data normalization calculation

Table 5 shows the results of the tourism destination rank experiment. Fuzzy AHP and AHP calculation results compared to the manual calculation to check the accuracy methods. Based on the MSE equation (5), Fuzzy AHP and AHP have an average value MSE of all data is ≈ 0.0002. The MSE results show that these methods contribute to the rank of the destinations in East Asia. This is because the MSE value generated is minimal.

IV. CONCLUSION

The implementation of AHP and Fuzzy AHP, when applied to the ranking of tourist attractions in East Asia, can prioritize and rank according to the actual calculated data. The accuracy of AHP and Fuzzy AHP in this experiment can produce the rank of the tourism trends in East Asia. Based on the experiments, these methods successfully rank with an MSE value of 0.0002 for calculation data based on actual calculations and normalization results. Based on the experiment, the Fuzzy-AHP process is more straightforward than AHP and accessible than the manual.

ACKNOWLEDGEMENT
The author is grateful to LPPM UPN "Veteran" Yogyakarta for its support in this scientific publication. In addition, the authors also thank the Department of Informatics for the support and assignment to carry out scientific publications in one of the journals.